\documentstyle[aps,pra,twocolumn]{revtex}
\RequirePackage{epsfig}
\def\L{{\cal L}}
\def\H{{\cal H}}
\def\U{\mbox{U}}
\def\vec#1{{\bf #1}}
\def\op#1{\hat{#1}}
\def\ket#1{| #1 \rangle}
\def\bra#1{\langle #1 |}
\def\lket#1{| #1 \rangle\rangle}
\def\lbra#1{\langle\langle #1 |}
\def\ave#1{\langle #1 \rangle}
\def\norm#1{\left|\!\left| #1 \right|\!\right|}
\def\ip#1#2{\langle #1 \mid  #2 \rangle}
\def\lip#1#2{\langle\langle #1 \mid  #2 \rangle\rangle}
\def\trace#1{\mathop{\rm Tr}\nolimits \left( #1 \right)} 

\newcounter {math}

\def\Bs{\!\!\!}
\begin{document}
\bibliographystyle{prsty}
%\draft
%\preprint{}
\title{Efficient Algorithm for Optimal Control of Mixed-State Quantum Systems}
\author{S.~G.~Schirmer}
\address{Department of Mathematics and Institute of Theoretical Science, 
         University of Oregon, Eugene, Oregon 97403}
\author{M.~D.~Girardeau}
\address{Department of Physics and Institutes of Theoretical Science and 
         Chemical Physics, \\
         University of Oregon, Eugene, Oregon 97403}
\author{J.~V.~Leahy}
\address{Department of Mathematics and Institute of Theoretical Science, 
         University of Oregon, Eugene, Oregon 97403}
\date{\today} 
\maketitle
\begin{abstract}
  In \cite{98ZR} Zhu and Rabitz presented a rapidly convergent iterative
  algorithm for optimal control of the expectation value of a positive definite
  observable in a pure-state quantum system.  In this paper we generalize this 
  algorithm to a quantum statistical mechanics setting and show that it is both
  efficient in the mixed-state case and effective in achieving the control
  objective of maximizing the ensemble average of arbitrary observables in the
  cases studied.
\end{abstract}
 %Insert valid PACS numbers
\narrowtext
\section{Introduction}
\label{sec:Intro}
Much work has recently been done on control of pure-state quantum systems using
the traditional wave-function formalism \cite{98ZR,99ZR,99ZSR}.  This work is
most important; however many physical systems, such as systems initially in 
thermal equilibrium or otherwise described by an ensemble of states, or systems
where dissipative processes are significant, can not be treated using this 
approach.  Therefore, a development of optimal control for mixed-state quantum 
systems is necessary.  In this paper we shall focus on generalizing an efficient 
iterative algorithm for quantum control \cite{98ZR} to a quantum statistical 
mechanics setting used in previous work \cite{98GSLK,97GISG,95KG,94GP,93YGWWM}.
This work is closely related to recently published, independently developed 
work by Yukiyoshi, Zhu and Rabitz \cite{99OZR} on quantum optimal control for 
systems with dissipation.  However, in our work we do not consider dissipation
terms since those terms are represented by non-Hermitian operators resulting in
non-unitary evolution of the system.  Unfortunately, the very accurate numerical 
implementation of the algorithm we propose depends on unitary evolution, as do 
the results on kinematical bounds \cite{98GSLK} and controllability \cite{95RSDRP},
which we use to show that the actual global maximum is reached by this algorithm.
\section{Mathematical Setup}
\label{sec:MathSetup}
As in our previous work, we consider a quantum-mechanical system whose state 
space $\H$ is a separable Hilbert space.  Any mixed state of the system can be
represented by a density operator $\op{\rho}(t)$ (acting on $\H$) with
eigenvalue decomposition
\begin{equation}
   \op{\rho}(t) = \sum_k w_k \ket{\Psi_k(t)}\bra{\Psi_k(t)},
\end{equation}
where $w_k$ are the eigenvalues, and $\ket{\Psi_k(t)}$ the corresponding
normalized eigenstates of $\op{\rho}(t)$, which evolve in time according 
to the time-dependent Schr{\"o}dinger equation. The eigenvalues satisfy
\begin{equation}
   0 \le w_k \le 1 \quad \forall k \mbox{ and } \sum_k w_k =1,
\end{equation}
i.e., they can be ordered in a (possibly finite) non-increasing sequence 
\[
 w_1 \ge w_2 \ge \ldots \ge w_k \ge \ldots \ge 0.
\]
Unless otherwise mentioned, the word state will in the following refer to a 
mixed state represented by a density operator $\op{\rho}(t)$.

The dynamical law for the system is given by the \emph{quantum Liouville 
equation}
\begin{equation} \label{Eq:CQL}
 \frac{\partial}{\partial t}\op{\rho}(t)=-\frac{i}{\hbar} [\op{H},\op{\rho}(t)].
\end{equation}
where $\op{H}$ is the (total) Hamiltonian of the system and $\op{\rho}(t_0)=
\op{\rho}_0$ defines the \emph{initial state} of the system (at time $t=t_0$).

Observables are represented by Hermitian operators $\op{A}$ on $\H$ and we 
define their expectation value to be the \emph{ensemble average} 
\begin{equation}
     \ave{\op{A}(t)} = \trace{\op{A} \op{\rho}(t)}.
\end{equation}

The set of bounded linear operators $\op{A}$ on $\H$ forms itself a Hilbert 
space, usually called \emph{Liouville space} and it is convenient to assign to 
each operator $\op{A}$ (on $\H$) a Liouville ket $\lket{A}$ denoting its 
representation in Liouville space.  The dual of $\lket{A}$ will be denoted by 
the Liouville bra $\lbra{A}$.  The inner product in Liouville space is defined 
by
\begin{equation}
  \lip{A}{B} = \trace{\op{A}^\dagger \op{B}}.
\end{equation}

Thus, an arbitrary mixed state of the system is represented by a Liouville ket 
$\lket{\rho(t)}$ that satisfies
\begin{equation} 
    \frac{\partial}{\partial t} \lket{\rho(t)} 
 = -\frac{i}{\hbar} {\cal L}(t) \lket{\rho(t)}
\end{equation}
with some initial condition $\lket{\rho(t_0)}=\lket{\rho_0}$.  ${\cal L}$ is the
\emph{Liouville operator} defined by the dual correspondence
\begin{equation}
   {\cal L} \lket{\rho(t)} \leftrightarrow [\op{H},\op{\rho}(t)].
\end{equation}
The expectation value $\ave{\op{A}(t)}$ of the observable $\op{A}$ is given by the
Liouville inner product $\lip{A}{\rho(t)}$.

\section{Controlling the Dynamics}
\label{sec:Control}

If the number $M$ of external \emph{control functions} 
\begin{equation}
  \vec{f}(t)=(f_1(t),f_2(t),\ldots,f_M(t)).
\end{equation}
acting on the system is finite and the system is \emph{control-linear} then
the total Hamiltonian of the system can be decomposed as follows:
\begin{equation} \label{Eq:Hamiltonian}
  \op{H}=\op{H}_0+\sum_{m=1}^M f_m(t)\op{H}_m.
\end{equation}
In this case, the corresponding Liouville operator also decomposes:
\begin{equation}
  {\cal L}={\cal L}_0+\sum_{m=1}^M f_m(t){\cal L}_m.
\end{equation}
The restrictions imposed on the controls depend on the particular system studied. 
However, a reasonable minimal requirement for the control functions $f_m(t)$ is 
that they should be bounded, measurable, real-valued functions defined on a time
interval $[t_0,t_F]$ that depends on the application.  

In the remainder of this paper we shall furthermore assume that there is only one 
control $f(t)$ acting on the system, which is sufficient for many applications of 
laser control.  However, we would like to point out that it is possible to 
generalize the algorithm to the case where there are multiple controls, such as 
two laser fields with perpendicular polarization driving the system.

Our goal is to maximize the expectation value (ensemble average) of a given 
observable, e.g., the population of a particular energy level or subspace of 
quantum states, the energy of a molecular bond, etc., at some fixed target time 
$t=t_F$ subject to certain constraints.  

More precisely, we define a functional \cite{94GP,95KG,81BV}
\begin{equation} \label{Eq:W}
  W(f,\rho_v,A_v) = W_1(\rho_v)-W_2(f,\rho_v,A_v)-W_3(f),
\end{equation}
whose value at a certain target time $t_F$ we would like to maximize.  $W_1$ is 
the expectation value of $\op{A}$ which we wish to maximize at the target time 
$t_F$,
\begin{equation} \label{Eq:WI}
  W_1(f) = \ave{A(t_F)} = \lip{A}{\rho_v(t_F)};
\end{equation}
$W_2$ and $W_3$ are constraint functionals, which we define as follows:
\begin{equation} \label{Eq:WII}
 W_2(f,\rho_v,A_v)
=\int_{t_0}^{t_F}\lbra{A_v(t)}\mbox{\(\frac{\partial}{\partial t}+\frac{i}{\hbar}\)}\L(t)\lket{\rho_v(t)} \; dt,
\end{equation}
\begin{equation} \label{Eq:WIII}
  W_3(f) = \frac{\lambda}{2} \int_{t_0}^{t_F} \!\!\! f^2(t) \; dt.
\end{equation}
$W_2$ ensures that the quantum Liouville equation is satisfied.  $W_3$ constrains 
the fluence, i.e., the total energy of the pulse.

$\rho_v(t)$ and $A_v(t)$ are variational trial functions that must satisfy the 
boundary conditions
\begin{equation} \label{Eq:bc}
   \rho_v(t_0)= \rho(t_0)=\rho_0, \quad A_v(t_F) = A.
\end{equation}
For simplicity we shall in the following choose units such that $\hbar=1$ and 
define $\partial_t = \frac{\partial}{\partial t}$.

Eqs (\ref{Eq:WI})--(\ref{Eq:bc}) are the generalization to Liouville space of the
Hilbert space formulation in \cite{98ZR}.  The details of the connection with
this paper will be discussed in appendix \ref{sec:rabitz}.

The solution of this control problem requires finding an admissible control 
$\vec{f}(t)$ such that $W$ and thus $\ave{\op{A}(t)}$ will attain its global 
maximum at time $t=t_F$.

\section{Algorithm}
\label{sec:algorithm}

We start by guessing an initial control $f^{(0)}(t)$ and determining an 
initial $\lket{\rho_v^{(0)}(t)}$ by solving
\[ 
  \partial_t\lket{\rho_v^{(0)}(t)}
 =-i\left[\L_0+f^{(0)}(t)\L_1\right]\lket{\rho_v^{(0)}(t)} 
\]
with initial condition $\lket{\rho_v^{(0)}(t_0)}=\lket{\rho_0}$.

For $n\ge 1$ and $k=0,1$ we define
\begin{equation}
  f^{(n,k)}(t) \equiv 
 -\frac{i}{\lambda}\lbra{A_v^{(n)}(t)}\L_1\lket{\rho_v^{(n-k)}(t)}
\end{equation}
\begin{equation}
 \L^{(n,k)}(t) \equiv \L_0 + f^{(n,k)}(t) \L_1
\end{equation}
and solve iteratively
\begin{eqnarray}
{\partial_t \lket{A_v^{(n)}(t)}}   &=& {-i\L^{(n,1)}(t)\lket{A_v^{(n)}(t)}}
\label{Eq:na}\\
{\partial_t \lket{\rho_v^{(n)}(t)}}&=& {-i\L^{(n,0)}(t)\lket{\rho_v^{(n)}(t)}}
\label{Eq:nb}
\end{eqnarray}
with the boundary conditions
\[ 
 \lket{A_v^{(n)}(t_F)} = \lket{A}, \quad
 \lket{\rho_0^{(n)}(t_0)} = \lket{\rho_0}.
\]
We observe that $f^{(n,k)}(t)$ is real.  Hence $\L^{(n,k)}$ is Hermitian and the
time-evolution of both $\ket{A_v^{(n)}(t)}$ and $\ket{\rho_v^{(n)}(t)}$ is unitary,
i.e.,
\begin{equation} \label{Eq:BdI}
  \norm{A_v^{(n)}(t)}_2 = \norm{A}_2 \mbox{ and }
  \norm{\rho_v^{(n)}(t)}_2 = \norm{\rho_0}_2
\end{equation}
for all $t\in [t_0,t_F]$ and any $n$. Furthermore,
\begin{equation}\label{Eq:normrho}
  \norm{\rho_0}_2^2=\trace{\op{\rho}_0^\dagger \op{\rho}_0}=\trace{\rho_0^2} \le 1.
\end{equation}
This algorithm can be shown to converge quadratically and monotonically as does
the pure-state version due to Zhu and Rabitz.  The details of the proof can be 
found in appendix \ref{sec:conv}.  However, we have no guarantee that $W_1(f)$
indeed assumes its \emph{global} maximum for this $f(t)$.  Additional criteria,
such as kinematical bounds and knowledge about controllability of the system are
necessary to decide if the control the algorithm produced is indeed optimal in 
the sense of steering the system to a global maximum of $W_1(f)$.

\section{Numerical Implementation}
\label{sec:numerics}

The differential equations arising from this feedback algorithm must be solved 
numerically.  While there are many methods of integrating differential equations 
numerically, we employ a symmetric split operator method \cite{98ZR,90S}.  The 
main advantage of this method is that it preserves the norm of the operators 
involved, which is of great importance in this problem.

We divide the time interval $[t_0,t_F]$ in subintervals $[t_j,t_{j+1}]$ of a fixed 
length $\Delta t=t_{j+1}-t_j$.  On each subinterval $[t_j,t_{j+1}]$ we approximate 
$f^{(n,k)}(t)$ by the constant $f^{(n,k)}(\tau_j)$ where 
\begin{equation}
  \tau_j = t_j + \Delta t/2 = t_{j+1}-\Delta t/2.
\end{equation}
With this approximation the propagator can be written as
\begin{equation} \label{Eq:Unk-exact}
   \U^{(n,k)}(t_{j+1},t_j) = \exp(-i\Delta t (\L_0 + f^{(n,k)}(\tau_j) \L_1)).
\end{equation}

For arbitrary matrices $A$ and $B$ we have 
\[ 
  e^{-i \alpha (A+B)} = e^{-i (\alpha/2) A} e^{-i \alpha B} e^{-i (\alpha/2) B}.
\]
up to second order terms in $A$ and $B$.  Thus (\ref{Eq:Unk-exact}) agrees to second 
order with
\begin{equation}
   e^{-i \frac{\Delta t}{2} \L_0}
   e^{-i\Delta t f^{(n,k)}(\tau_j) \L_1}
   e^{-i \frac{\Delta t}{2} \L_0}.
\end{equation}

This symmetric splitting is numerically favorable since it allows us to reduce the 
matrix exponentials to a simple linear combination of complex exponentials:
\begin{eqnarray}
 {\U_0} &\equiv& {\exp(-i\Delta t\L_0/2)}\nonumber\\
 &=& {\sum_{a=1}^N \lket{a} e^{-i a \Delta t/2} \lbra{a}}\\
 {\U_1^{(n,k)}(\tau_j)} &\equiv& {\exp(-i \Delta t f^{(n,k)}(\tau_j) \L_1)}
 \nonumber\\
 &=& {\sum_{b=1}^N \lket{b} e^{-i \Delta t f^{(n,k)}(\tau_j)b}\lbra{b}} 
\end{eqnarray}
where $\lket{a}$ and $\lket{b}$ are the eigenkets of $\L_0$ and $\L_1$, respectively; $a$ 
and $b$ are the corresponding (real) eigenvalues.  This leads to
\begin{equation} \label{Eq:Unk}
 \U^{(n,k)}(\tau_j)
 \equiv \sum_{a,b=1}^N \! |\lip{a}{b}|^2 e^{-i\Delta t(a+b f^{(n,k)}(\tau_j))}
 \lket{a}\lbra{a}.
\end{equation}
$\U^{(n,k)}(\tau_j)$ agrees up to second order with $\U^{(n,k)}(t_{j-1},t_j)$.  Since 
$\L_0$ and $\L_1$ do not depend on $f^{(n,k)}$, the eigenvalue decomposition needs to be 
done only once, i.e., the only quantities that need to be computed in each step of the 
iteration are the complex exponentials $e^{-i\Delta t(a+b f^{(n,k)}(\tau_j))}$ for 
all possible values of $a$ and $b$.

In order to compute $f(\tau_j)$, we note that 
\begin{equation}
 f(t \pm\Delta t) \approx f(t) \pm \Delta t \frac{df}{dt} (t)
\end{equation}
to 1st order, and hence we have
\begin{eqnarray}
{f^{(n,0)}(\tau_j)} 
&=&{f^{(n,0)}(t_j)}\nonumber\\
& &{+\frac{\Delta t}{2\lambda}\lip{A_v^{(n)}(t_j)}{[\L_0,\L_1] \rho_v^{(n)}(t_j)}}
\label{Eq:ftauO}\\
{f^{(n,1)}(\tau_{j-1})} 
&=&{f^{(n,1)}(t_j)}\nonumber\\
& &{-\frac{\Delta t}{2\lambda}\lip{A_v^{(n)}(t_j)}{[\L_0,\L_1] \rho_v^{(n-1)}(t_j)}.}
\label{Eq:ftauI}
\end{eqnarray}

\section{Illustrative Computations}
\label{sec:illus}

As an example for molecular quantum control, we consider a Morse oscillator model
for a diatomic molecule with $N$ discrete energy levels $E_n$ corresponding to 
independent vibrational eigenstates $\ket{n}$ of the system. The unperturbed 
Hamiltonian is thus
\begin{equation}
  \op{H}_0 = \sum_{n=1}^N E_n \ket{n} \bra{n}.
\end{equation}
The interaction Hamiltonian of the driven system can be approximated by 
$\op{H}_1=f(t)\op{V}$ where $f(t)$ is an external laser field that serves as 
control function, and $\op{V}$ is the transition operator, which we choose to be 
of the dipole form
\begin{equation}
  \op{V}=\sum_{n=1}^{N-1} d_n (\ket{n} \bra{n+1} + \ket{n+1}\bra{n}).
\end{equation}
This system is completely controllable, which can easily be verified using an
algorithm described in \cite{95RSDRP}.  Thus, the global minima and maxima of any 
observable are determined by the kinematical bounds and these extrema are 
dynamically attainable.

For the sake of illustration we choose $N=4$.  The corresponding energy 
levels are $E_1=0.4843$, $E_2=1.4214$, $E_3=2.3691$ and $E_4=3.2434$ in units of 
$\hbar\omega_0$ where $\omega_0=7.8\times 10^{14} \mbox{ s}^{-1}$ for HF.

Let us first assume that the system is initially in the ground state, i.e., 
$\op{\rho}_0=\ket{1}\bra{1}$ and that our goal is to maximize the vibrational 
energy of the bond, i.e., $\op{A}=\op{H}_0$. In this case, the results on 
kinematical bounds in \cite{98GSLK} give
\begin{equation}
  1.4214 \le \ave{\op{A}(t)} \le 3.2434.
\end{equation}
The lower bound is attained exactly if the population of level 1 (ground state) 
is 1.  The upper bound is attained exactly if the population of level 4 (highest 
state) is 1. Figs 1-3 show the results of our computations using the algorithm
described above.  Starting with a randomly generated function $f$ of sufficiently 
small magnitude and $\lambda=4$, the observable rapidly approaches its converged
value within only a few iterations.  Fig.~1 shows the final pulse $f(t)$, Fig.~2 
the corresponding evolution of the populations of energy levels 1 through 4, and 
Fig.~3 shows the evolution of the expectation value of the observable.  At the 
target time $t_F=200$ fs, we observe a nearly complete inversion of the populations, 
with the population of level four being close to 97\%.  $\ave{\op{A}(t_F)}$ is about 
98\% of the theoretical maximum.

Secondly, we assume that the system is initially in thermal equilibrium, i.e.,
\[
  \op{\rho}_0=\sum_{n=1}^N w_n \ket{n}\bra{n}
\] 
with weights 
\[
  w_n=C\exp(-E_n/(E_4-E_1)).
\]
This is a Bolzmann distribution with $kT=E_4-E_1$. 
\[
  C=(e^{-E_1/kT}+e^{-E_2/kT}+e^{-E_3/kT}+e^{-E_4/kT})^{-1}
\] 
is the normalization constant. Concretely, $w_1=0.3850$, $w_2=0.2758$, $w_3=0.1976$ 
and $w_4=0.1416$.  According to \cite{98GSLK},
\begin{equation}
  1.5059 \le \ave{\op{A}} \le 2.2592.
\end{equation}
The lower bound is attained in thermal equilibrium.  The upper bound is attained
exactly if the populations are inverted, i.e., the most energetic state (here 
$n=4$) has the highest population, the second most energetic state has the second 
highest population, etc.  Figs 4-6 show the results of our computations using the 
algorithm described above.  Again, we started with a randomly generated function $f$ 
of sufficiently small magnitude and $\lambda=4$.  Fig.~4 shows the final pulse $f(t)$,
Fig.~5 the corresponding evolution of the populations of energy levels 1 through 4, 
and Fig.~6 shows the evolution of the expectation value of the observable.  At the 
target time $t_F=200$ fs we observe a nearly complete inversion of the populations 
with $\ave{\op{A}(t_F)}$ being 99\% of the theoretical maximum.

\begin{figure} \label{FigI}
\epsfysize 2.5in
\epsfbox{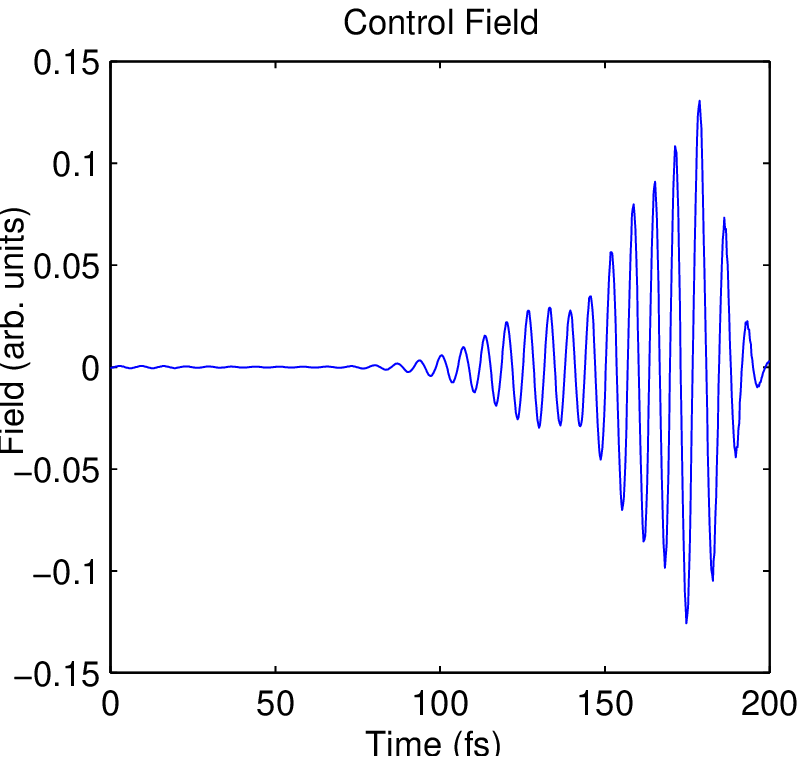}
\caption{Optimal pulse for a four-level Morse oscillator with $\op{\rho}_0=\ket{1}\bra{1}$}
\epsfysize 2.5in
\epsfbox{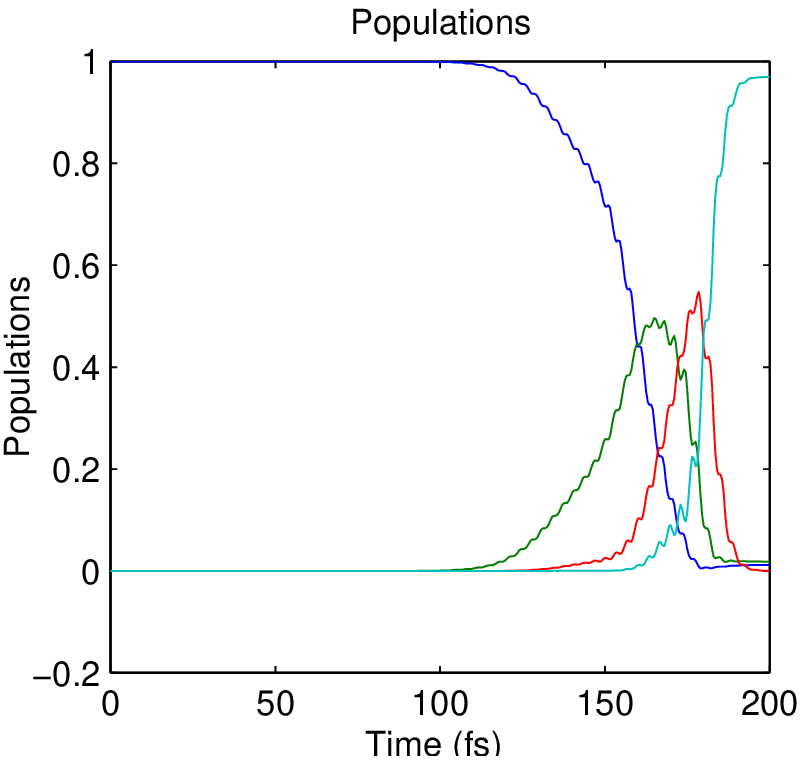}
\caption{Evolution of the populations for a four-level Morse oscillator with $\op{\rho}_0=\ket{1}\bra{1}$}

\epsfysize 2.5in
\epsfbox{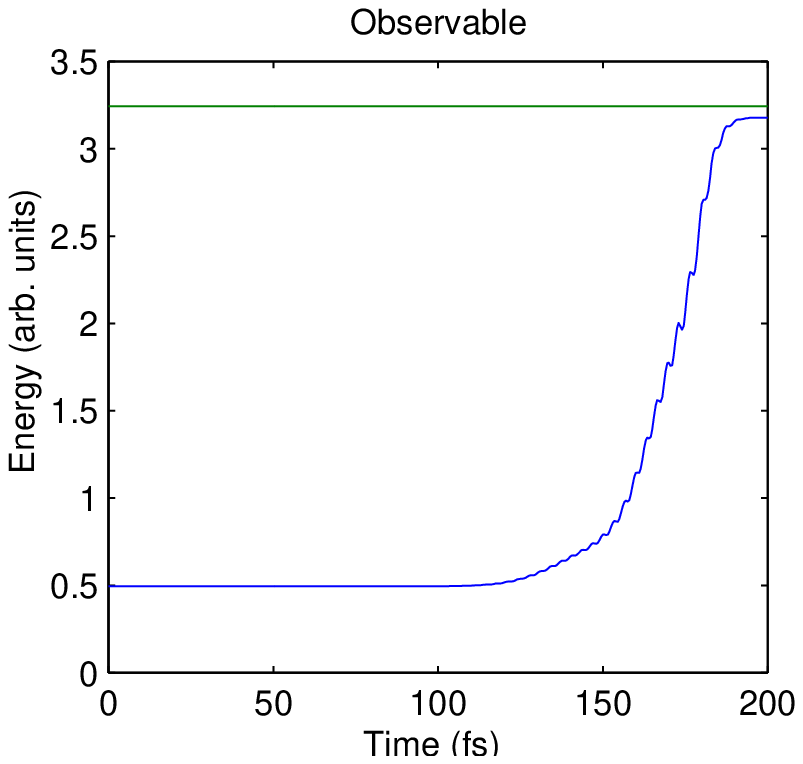}
\caption{Evolution of the vibrational energy for a four-level Morse oscillator with $\op{\rho}_0=\ket{1}\bra{1}$}
\end{figure}

\begin{figure} \label{FigII}
\epsfysize 2.5in
\epsfbox{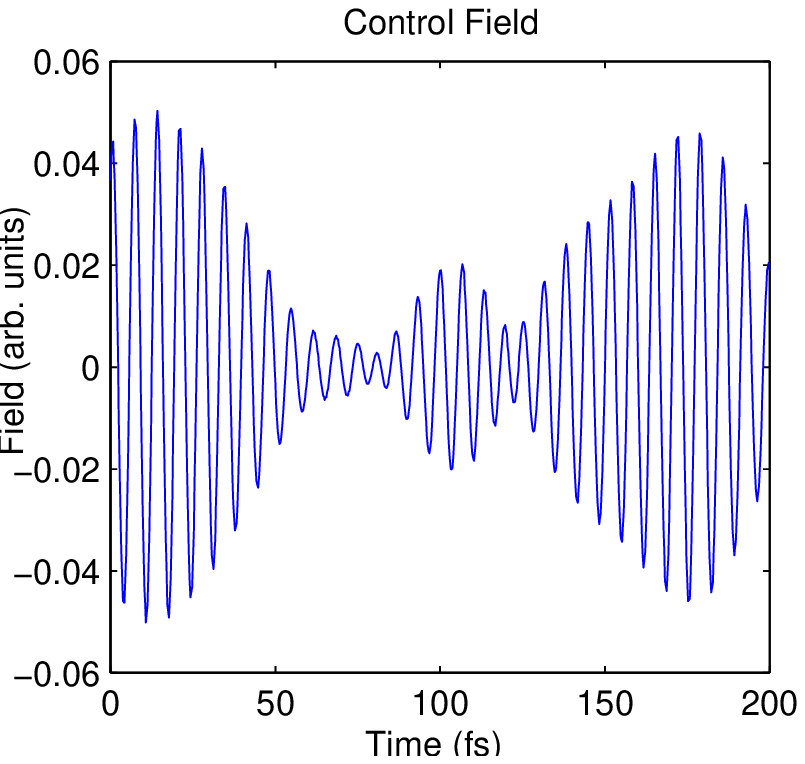}
\caption{Optimal pulse for a four-level Morse oscillator with $\op{\rho}_0=\sum_{n=1}^4\ket{n}\bra{n}$}

\epsfysize 2.5in
\epsfbox{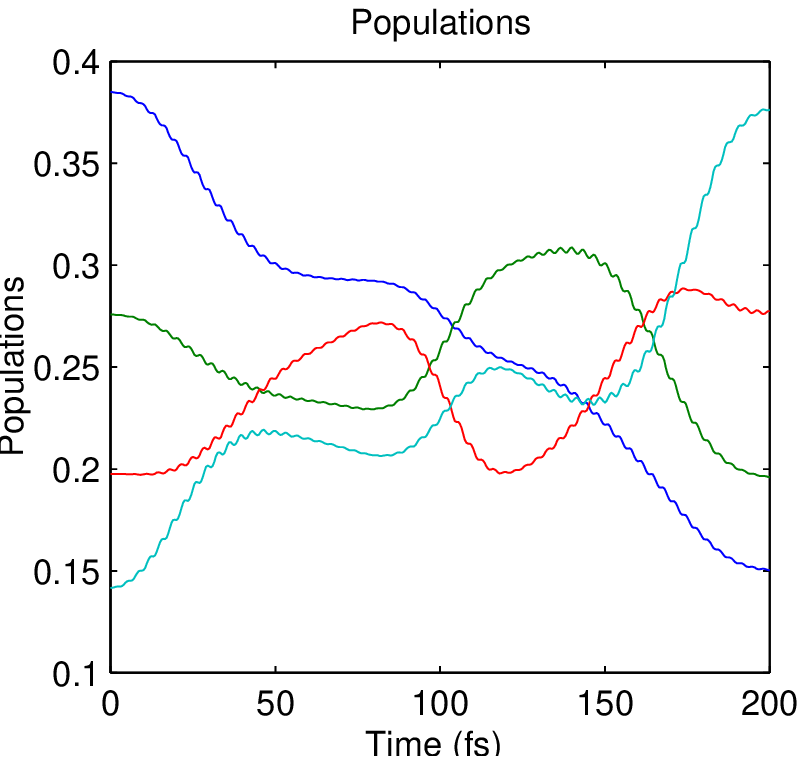}
\caption{Evolution of the populations for a four-level Morse oscillator with $\op{\rho}_0=\sum_{n=1}^4\ket{n}\bra{n}$}

\epsfysize 2.5in
\epsfbox{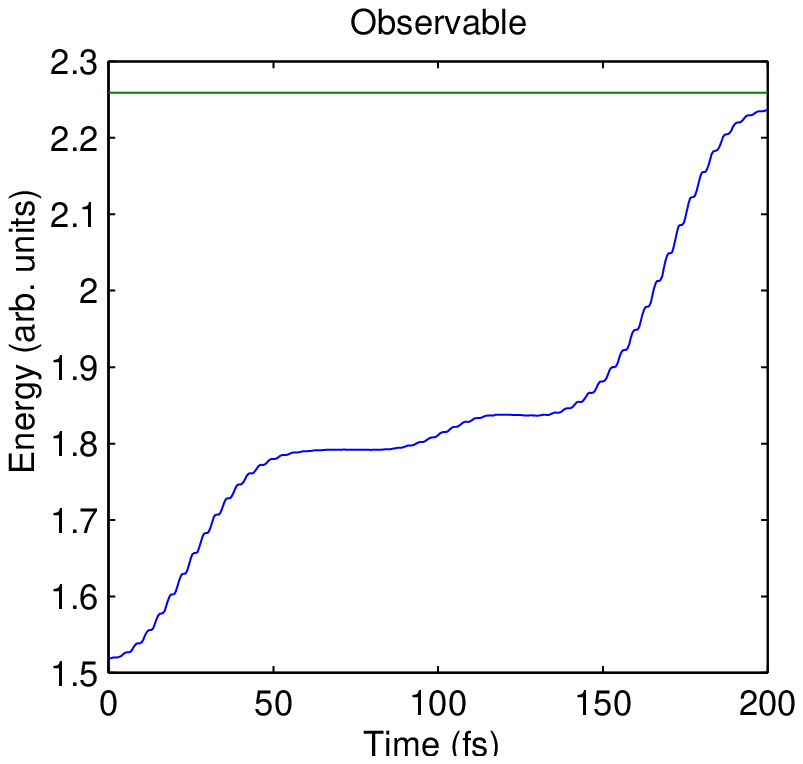}
\caption{Evolution of the vibrational energy for a four-level Morse oscillator with $\op{\rho}_0=\sum_{n=1}^4\ket{n}\bra{n}$}
\end{figure}
\section{Conclusion}
\label{sec:conclusion}
In this paper we demonstrated that an efficient algorithm for optimal control
of quantum systems can be applied in a quantum statistical mechanics setting and
that this algorithm is also highly effective in realizing the control objective 
of maximizing the ensemble average of an observable.
\section{Acknowledgements}
The authors would like to thank the referee for his helpful suggestions and for
pointing out the connection with \cite{99OZR}.
\appendix
\section{Relation to Work of Rabitz et al.}
\label{sec:rabitz}

Our variational functional and Euler-Lagrange equations are equivalent to the ones
used in \cite{98ZR} in the pure state limit, i.e., if $\op{\rho}_v(t)=
\ket{\psi_v(t)}\bra{\psi_v(t)}$ where $\ket{\psi_v(t)}$ is a normalized state then
\begin{eqnarray}
 {W}&=&{\bra{\psi_v(t_F)}A\ket{\psi_v(t_F)}-\alpha_0 \int_{t_0}^{t_F} f^2(t)dt}
\nonumber\\
 &&{-2\Re\int_{t_0}^{t_F}\bra{\chi_v(t)}[\partial_t+i\op{H}(f,t)]\ket{\phi_v(t)}dt}
\end{eqnarray}

Choose a (time-dependent) complete orthonormal set $\{\ket{\psi_n(t)}:n=1,2,
\ldots\}$ such that $\ket{\psi_1(t)}=\ket{\psi_v(t)}$ for all $t$. Then we have
\begin{eqnarray*}
W_1 
&=& {\trace{\op{A}\op{\rho}_v(t_F)}}\\
&=& {\sum_n \bra{\psi_n(t_F)}\op{A}_v(t_F)\ket{\psi_v(t_F)}
     \ip{\psi_v(t_F)}{\psi_n(t_F)}}\\
&=& {\bra{\psi_v(t_F)}A\ket{\psi_v(t_F)}.}
\end{eqnarray*}
Furthermore, setting $\ket{\chi_v(t)} = \op{A}_v(t) \ket{\psi_v(t)}$ we obtain
\begin{eqnarray*}
&&{\lip{A_v(t)}{\partial_t \rho_v(t)}
  =\trace{\op{A}_v(t)\partial_t \op{\rho}_v(t)}}\\
&=&{\sum_n\bra{\psi_n(t)}\op{A}_v(t)(\partial_t\ket{\psi_v(t)})
    \ip{\psi_v(t)}{\psi_n(t)}}\\
&&{+\sum\bra{\psi_n(t)}\op{A}_v(t)\ket{\psi_v(t)}
    (\partial_t\bra{\psi_v(t)})\ket{\psi_n(t)}}\\
&=&{\bra{\psi_v(t)}\op{A}_v(t)\partial_t\ket{\psi_v(t)}}\\
&&{+\sum_n (\partial_t\bra{\psi_v(t)})\ket{\psi_n(t)}\bra{\psi_n(t)}
    \op{A}_v(t)\ket{\psi_v(t)}}\\
&=&{\bra{\psi_v(t)}\op{A}_v(t)\partial_t\ket{\psi_v(t)}
    +(\partial_t\bra{\psi_v(t)})\op{A}_v(t)\ket{\psi_v(t)}}\\
&=&{\bra{\psi_v(t)}\op{A}_v(t)\partial_t\ket{\psi_v(t)}
    +(\bra{\psi_v(t)}\op{A}_v(t)\partial_t\ket{\psi_v(t)})^*}\\
&=&{2\Re \bra{\psi_v(t)}\op{A}_v(t)\partial_t\ket{\psi_v(t)}}\\
&&{=2\Re \ip{\chi_v(t)}{\partial_t\psi_v(t)}}
\end{eqnarray*}
and
\[ \begin{array}{rcl}
& &{\lip{A_v(t)}{i\L(f,t)\rho_v(t)}}\\
&=&{i\trace{\op{A}_v(t)[\op{H}(f,t),\op{\rho}_v(t)]}}\\
&=&{\sum_n i\bra{\psi_n(t)}\op{A}_v(t)\op{H}(f,t)\ket{\phi_v(t)}
    \ip{\psi_v(t)}{\psi_n(t)}}\\
&&{-\sum_n i\bra{\psi_n(t)}\op{A}_v(t)\ket{\phi_v(t)}\bra{\psi_v(t)}
   \op{H}(f,t)\ket{\psi_n(t)}}\\
&=&{i\bra{\psi_v(t)}\op{A}_v(t)\op{H}(f,t)\ket{\phi_v(t)}}\\
&&{-\sum_n i\bra{\psi_v(t)}\op{H}(f,t)\ket{\psi_n(t)}\bra{\psi_n(t)}\op{A}_v(t)\ket{\phi_v(t)}}\\
&=&{\bra{\chi_v(t)}i\op{H}(f,t)\ket{\phi_v(t)}-\bra{\psi_v(t)}i\op{H}(f,t)\ket{\chi_v(t)}}\\
&=&{\bra{\chi_v(t)}i\op{H}(f,t)\ket{\phi_v(t)}+(\bra{\chi_v(t)}i\op{H}(f,t)\ket{\phi_v(t)})^*}\\
&=&{2\Re \bra{\chi_v(t)}i\op{H}(f,t)\ket{\phi_v(t)}.}
\end{array}\]
Hence, we have
\begin{eqnarray*}
W_2 &=& {\int_{t_0}^{t_F}\lbra{A_v(t)}\partial_t+i\L(t)\lket{\rho_v(t)}dt}\\
    &=& {2\Re \int_{t_0}^{t_F}\bra{\chi_v(t)}[\partial_t+i\op{H}(f,t)]
         \ket{\phi_v(t)}dt}
\end{eqnarray*}
in the pure state case. $W_3$ remains essentially the same, i.e., we simply set
$\alpha_0=\lambda/2$.  The equivalence of the Euler-Lagrange equations follows.

\section{Proof of Convergence Properties}
\label{sec:conv}
After the $n$th iteration step, the objective functional is
\begin{eqnarray}
 {W^{(n)}}
 &=&{W_1^{(n)}-W_3^{(n)}}\nonumber\\
 &=&{\lip{A}{\rho_v^{(n)}(t_F)}-\frac{\lambda}{2}\int_{t_0}^{t_F}[f^{(n,0)}(t)]^2 dt}
\end{eqnarray}
since $W_2^{(n)}=W_2(f^{(n,0)},\rho_v^{(n)},A_v^{(n)})=0$ according to Eqs 
(\ref{Eq:WII}) and (\ref{Eq:nb}).

{\bf Lemma:}
$W^{(n)}$ is \emph{uniformly bounded.}

\begin{proof}{}
Cauchy-Schwarz's inequality and Eqs (\ref{Eq:BdI}), (\ref{Eq:normrho}) give
\[
|\lip{A}{\rho_v^{(n)}(t)}|^2\le\norm{A}_2^2\cdot\norm{\rho_v^{(n)}(t)}_2^2\le\norm{A}_2^2,
\]
as well as
\begin{eqnarray*}
{|f(t)|^2}
& = &{\left|-\frac{i}{\lambda} \lip{A_v^{(n)}(t)}{\L_1 \rho_v^{(n)}(t)} \right|^2}\\
&\le&{\frac{1}{\lambda^2} \norm{A_v^{(n)}(t)}_2^2 \cdot \norm{\L_1 \rho_v^{(n)}(t)}_2^2}\\
&\le&{\frac{1}{\lambda^2} \norm{A_v^{(n)}(t)}_2^2 \cdot \norm{\L_1} \cdot 
       \norm{\rho_v^{(n)}(t)}_2^2}\\
&\le&{\frac{1}{\lambda^2} \norm{A_v^{(n)}(t)}_2^2 \cdot \norm{\L_1}}
\end{eqnarray*}
where $\norm{\L_1}$ is the usual operator norm.  Thus, 
\begin{eqnarray*}
 {|W^{(n)}|} 
 &\le& {|W_1^{(1)}| + |W_3^{(n)}|}\\
 &\le& {\norm{A}_2 + \frac{t_F-t_0}{2\lambda} \norm{A_v^{(n)}(t)}_2^2 \cdot \norm{\L_1}}
\end{eqnarray*}
for all $n$, which establishes the claim.
\end{proof}

{\bf Lemma:} If $\U(t,t_0)$ satisfies 
\[
  \partial_t \U(t,t_0) = -i \L(t) \U(t,t_0)
\]
then 
\[
 \lket{\rho(t)}=\U(t,t_0) \int_{t_0}^t \U^\dagger(t',t_0) \lket{\phi(t')} dt'
\]
is a solution of
\[
   \partial_t \lket{\rho(t)} = -i \L(t) \lket{\rho(t)} + \lket{\phi(t)}
\]
\begin{proof}{} Using the product rule and
\[
 \partial_t \int_{t_0}^t \U^\dagger(t',t_0) \lket{\phi(t')} dt'
= \U^\dagger(t,t_0) \lket{\phi(t)}.
\]
to differentiate $\lket{\rho(t)}$ gives
\begin{eqnarray*} \textstyle
{\partial_t \lket{\rho(t)}}
&=& {[-i \L(t) \U(t,t_0)] \int_{t_0}^t \U^\dagger(t',t_0) \lket{\phi(t')} dt'}\\
& & {+\U(t,t_0) \U^\dagger(t,t_0) \lket{\phi(t)}} \\
&=& {-i \L(t) \lket{\rho(t)} + \lket{\phi(t)}.} 
\end{eqnarray*}
\end{proof}
\begin{theorem}{\em Convergence.}
The sequence $\{W^{(n)}\}$ converges monotonically and quadratically in the 
control, i.e.,
\begin{eqnarray}
{0} &=& {\lim_{n\rightarrow\infty} W^{(n+1)} - W^{(n)}}\nonumber\\
    &=& {\lim_{n\rightarrow\infty} \frac{\lambda}{2}\int_{t_0}^{t_F}
         [\delta f^{(n+1)}(t)]^2+[\delta f^{(n+1,n)}(t)]^2\, dt}
\end{eqnarray}
\end{theorem}
\begin{proof}{} Setting
\begin{equation}
 \lket{\delta \rho_v^{(n)}(t)}=\lket{\rho_v^{(n+1)}(t)}-\lket{\rho_v^{(n)}(t)},
\end{equation}
\begin{eqnarray}
{\delta W^{(n+1,n)}} 
&=& {W^{(n+1)}-W^{(n)}}\nonumber\\
&=& {\lip{A}{\delta\rho_v^{(n)}(t_F)}} \nonumber \\
& & {-\frac{\lambda}{2} \int_{t_0}^{t_F} [f^{(n+1,0)}(t)]^2-[f^{(n,0)}]^2 dt.}
\nonumber\\
\end{eqnarray}
During the iteration
\begin{equation}
 \partial_t \lket{\rho_v^{(n)}(t)} = -i [\L_0 + f^{(n,0)}(t) \L_1] \lket{\rho_v^{(n)}(t)}.
\end{equation}
Hence, setting
\begin{eqnarray}
 {\delta f^{(n+1,n)}} &=& {f^{(n+1,1)}(t)-f^{(n,0)}(t)}\\
 {\delta f^{(n)}}     &=& {f^{(n,0)}(t)  -f^{(n,1)}(t)}
\end{eqnarray}
and noting that
\begin{eqnarray*}
& &{\L_1 f^{(n+1,1)} \lket{\delta \rho_v^{(n)}(t)}}\\
& &{+\L_1 \lket{\delta f^{(n+1)} \rho_v^{(n+1)}(t)+\delta f^{(n+1,n)}\rho_v^{(n)}(t)}}\\
&=&{\L_1 f^{(n+1,1)} \lket{\rho_v^{(n+1)}(t)}-\L_1 f^{(n+1,1)} \lket{\rho_v^{(n)}(t)}}\\
& &{+\L_1 f^{(n+1,0)} \lket{\rho_v^{(n+1)}(t)}-\L_1 f^{(n+1,1)}\lket{\rho_v^{(n+1)}(t)}}\\
& &{+\L_1 f^{(n+1,1)} \lket{\rho_v^{(n)}(t)}-\L_1 f^{(n,0)}\lket{\rho_v^{(n)}(t)}}\\
&=&{\L_1 f^{(n+1,0)} \lket{\rho_v^{(n+1)}(t)} 
   -\L_1 f^{(n,0)}   \lket{\rho_v^{(n)}(t)}}
\end{eqnarray*}
we obtain
\begin{eqnarray}
{\partial_t \lket{\delta \rho_v^{(n)}(t)}}
&=&{-i\L^{(n+1,1)} \lket{\delta \rho_v^{(n)}(t)}}\nonumber\\
& &{-i\L_1 \lket{(\delta f^{(n+1)} \rho_v^{(n+1)}+\delta f^{(n+1,n)}\rho_v^{(n)})(t)}}
\nonumber\\ \label{Eq:deltaRho}
\end{eqnarray}
Setting
\begin{equation}
  \U(t,t_0,f^{(n+1,1)})=\exp_+\left[-i\int_{t_0}^t\L^{(n+1,1)}(\tau) d\tau \right]
\end{equation}
where $\exp_+$ denotes the time-ordered exponential, the formal solution of 
\ref{Eq:deltaRho} is (according to the previous lemma) given by 
\begin{eqnarray}
{\lket{\delta \rho_v^{(n)}(t)}} 
&=& {-i\U(t,t_0,f^{(n+1,1)})\int_{t_0}^t\U^\dagger(t',t_0,f^{(n+1,1)}) \times}
\nonumber\\
&&{\L_1\lket{(\delta f^{(n+1)}\rho_v^{(n+1)}+\delta f^{(n+1,n)}\rho_v^{(n)})(t')}dt'.}
\nonumber\\
\end{eqnarray}
Observing that
\[
 \lket{A_v^{(n)}(t)} = \U(t,t_0,f^{(n+1,1)}) \U^\dagger(t_F,t_0,f^{(n+1,1)}) \lket{A}
\]
and thus
\[
 \lbra{A_v^{(n)}(t)} = \lbra{A}\U(t_F,t_0,f^{(n+1,1)}) \U^\dagger(t,t_0,f^{(n+1,1)}),
\]
we arrive at
\begin{eqnarray}
& &{\lip{A}{\delta \rho_v^{(n)}(t_F)}}\nonumber\\
&=&{-i\int_{t_0}^{t_F}\lbra{A}\U(t_F,t_0,f^{(n+1,1)})
                                       \U^\dagger(t,t_0,f^{(n+1,1)})\times} 
\nonumber\\
& &{\L_1\lket{(\delta f^{(n+1)}\rho_v^{(n+1)}+\delta f^{(n+1,n)}\rho_v^{(n)})(t)}dt.}
\nonumber\\
&=&{-i\int_{t_0}^{t_F}\lbra{A_v^{(n+1)}(t)}\times} \nonumber\\
& &{\L_1\lket{(\delta f^{(n+1)}\rho_v^{(n+1)}+\delta f^{(n+1,n)}\rho_v^{(n)})(t)}dt.}
\nonumber\\
&=&{\int_{t_0}^{t_F}\Bs-i\delta f^{(n+1)}(t)\lip{A_v^{(n+1)}(t)}{\L_1\rho_v^{(n+1)}(t)}dt}
\nonumber\\
&&{+\int_{t_0}^{t_F}\Bs-i\delta f^{(n+1,n)}(t)\lip{A_v^{(n+1)}(t)}{\L_1\rho_v^{(n)}(t)}dt}
\nonumber\\
&=&{\lambda\int_{t_0}^{t_F}\Bs\delta f^{(n+1)}(t) f^{(n+1,0)}(t)
                              +\delta f^{(n+1,n)}(t) f^{(n+1,1)}(t) dt}
\nonumber\\
&=& {\lambda\int_{t_0}^{t_F}\Bs [f^{(n+1,0)}(t)]^2-f^{(n+1,1)}(t) f^{(n+1,0)}(t)}
\nonumber\\
& &{+ [f^{(n+1,1)}(t)]^2 - f^{(n,0)}(t) f^{(n+1,1)}(t)dt}
\nonumber\\
\end{eqnarray}
\begin{eqnarray}
{\delta W^{(n+1,n)}}
&=&{\frac{\lambda}{2}\int_{t_0}^{t_F}\Bs[f^{(n+1,0)}(t)]^2+2f^{(n+1,1)}(t)f^{(n+1,0)}(t)}
\nonumber\\
& &{+2[f^{(n+1,1)}(t)]^2 - 2 f^{(n,0)}(t) f^{(n+1,1)}(t)}
\nonumber\\
& &{\qquad\qquad+[f^{(n,0)}(t)]^2 \, dt}
\nonumber\\
&=&{\frac{\lambda}{2}\int_{t_0}^{t_F}\Bs[\delta f^{(n+1)}(t)]^2+[\delta f^{(n+1,n)}(t)]^2dt}
\nonumber\\
\end{eqnarray}
and thus the total variation from $n=0$ to $n_F$ is
\begin{eqnarray}
 {\delta W^{(n_F,0)}}
 &=& {W^{(n_F)}-W^{(0)} =\sum_{n=0}^{n_F-1}\delta W^{(n+1,n)}} \nonumber\\
 &=& {\sum_{n=0}^{n_F-1} \frac{\lambda}{2}\int_{t_0}^{t_F}
                 [\delta f^{(n+1)}(t)]^2+[\delta f^{(n+1,n)}(t)]^2\, dt}\nonumber\\
\end{eqnarray}
Since $W^{(n)}$ is uniformly bounded, $W^{(n_F)}-W^{(0)}$ is also uniformly bounded for 
all $n_F$ and thus the sequence $\{\delta W^{(n_F,0)}:n_F\in\mbox{\bf N}_0\}$ is uniformly
bounded.
\[
  \frac{\lambda}{2}\int_{t_0}^{t_F}[\delta f^{(n+1)}(t)]^2+[\delta f^{(n+1,n)}(t)]^2 dt>0
\]
for any $n$ implies furthermore that $\delta W^{(n_F,0)}$ is an increasing sequence. 
Hence,
\[
  \lim_{n_F\rightarrow\infty} \delta W^{(n_F,0)} \mbox{ exists and is finite}.
\]
Consequently
\[
 \lim_{n\rightarrow\infty} \frac{\lambda}{2}\int_{t_0}^{t_F}
 [\delta f^{(n+1)}(t)]^2+[\delta f^{(n+1,n)}(t)]^2\, dt=0.
\]
\end{proof}
\bibliography{science}
\end{document}